\documentclass[journal]{IEEEtran}
\usepackage{cite}
\usepackage{amsmath,amssymb,amsfonts}
\usepackage{mathtools}
\usepackage{graphicx}
\usepackage{textcomp}
\usepackage{xcolor}
\usepackage{url}
\def\BibTeX{{\rm B\kern-.05em{\sc i\kern-.025em b}\kern-.08em
    T\kern-.1667em\lower.7ex\hbox{E}\kern-.125emX}}
\usepackage[utf8]{inputenc} 
\graphicspath{{../figures/}}
\usepackage[outdir=./converted_figures/]{epstopdf}  
\usepackage{makecell}  
\usepackage[
  separate-uncertainty = true,
  multi-part-units = repeat
]{siunitx}
\usepackage{bm}        
\usepackage{multirow}
\usepackage{upgreek}

\DeclarePairedDelimiter\norm{\lVert}{\rVert}%

\newcommand{\ivector}{intermediate switching vector }

\newcommand{\pmsm}{PMSM}

\newcommand{\udq}{\boldsymbol{u}_\mathrm{dq}}
\newcommand{\uidq}{\boldsymbol{u}_\mathrm{i,dq}}
\newcommand{\udqm}{\overline{\boldsymbol{u}}_\mathrm{dq}}
\newcommand{\uab}{\boldsymbol{u}_\mathrm{\alpha \beta}}

\newcommand{\idq}{\boldsymbol{i}_\mathrm{dq}}

\newcommand{\iabc}{\boldsymbol{i}_\mathrm{abc}}
\newcommand{\edq}{\boldsymbol{e}_\mathrm{dq}}

\newcommand{\idqref}{\boldsymbol{i}^*_{\mathrm{dq}}}

\newcommand{\Ldq}{\boldsymbol{L}_\mathrm{dq}}
\newcommand{\Ldqdiffm}{\boldsymbol{L}_\mathrm{dq,\Delta}}

\newcommand{\A}{\boldsymbol{A}}
\newcommand{\B}{\boldsymbol{B}}
\newcommand{\E}{\boldsymbol{E}}

\newcommand{\Tabcab}{\boldsymbol{T}_\mathrm{abc \upalpha \upbeta}}
\newcommand{\Tababc}{\boldsymbol{T}_\mathrm{ \upalpha \upbeta abc}}
\newcommand{\Tdqab}{\boldsymbol{T}_\mathrm{dq \upalpha \upbeta}}
\newcommand{\Tabdq}{\boldsymbol{T}_\mathrm{\upalpha \upbeta dq}}
\newcommand{\Tdqabc}{\boldsymbol{T}_\mathrm{dqabc}}
\newcommand{\Tabcdq}{\boldsymbol{T}_\mathrm{abcdq}}

\newcommand{\Psid}{{\psi_{\mathrm{d}}}}
\newcommand{\Psiq}{{\psi_{\mathrm{q}}}}
\newcommand{\Psidq}{{\bm{\psi}_{\mathrm{dq}}}}

\newcommand{\resd}{{r_{\mathrm{d}}}}
\newcommand{\resq}{{r_{\mathrm{q}}}}
\newcommand{\resdq}{{\boldsymbol{r}_{\mathrm{dq}}}}

\newcommand{\Itdd}{{I_{\mathrm{TDD}}}}

\newcommand{\Psip}{\psi_{\mathrm{p}}}
\newcommand{\Psipv}{\boldsymbol{\psi}_{\mathrm{p}}}
\newcommand{\UDC}{u_{\mathrm{DC}}}
\newcommand{\wel}{\omega}
\newcommand{\eel}{\varepsilon}
\newcommand{\ud}{u_\mathrm{d}}
\newcommand{\uq}{u_\mathrm{q}}
\newcommand{\udo}{\overline{u}_\mathrm{d}}
\newcommand{\uqo}{\overline{u}_\mathrm{q}}

\newcommand{\id}{i_\mathrm{d}}
\newcommand{\iq}{i_\mathrm{q}}
\newcommand{\rdv}{\boldsymbol{r}_\mathrm{d}}
\newcommand{\rqv}{\boldsymbol{r}_\mathrm{q}}

\newcommand{\Ld}{L_\mathrm{d}}
\newcommand{\Lq}{L_\mathrm{q}}
\newcommand{\Ldddiff}{L_\mathrm{dd}}
\newcommand{\Ldqdiff}{L_\mathrm{dq}}
\newcommand{\Lqddiff}{L_\mathrm{qd}}
\newcommand{\Lqqdiff}{L_\mathrm{qq}}
\newcommand{\Ts}{\mathrm{T_s}}
\newcommand{\Ti}{\mathrm{T_i}}
\newcommand{\Tr}{\mathrm{T}}

\newcommand{\Rs}{R_\mathrm{s}}

\newcommand{\sabc}{\boldsymbol{s}_{\mathrm{abc}}}
\newcommand{\siabc}{\boldsymbol{s}_{\mathrm{i,abc}}}

\usepackage{booktabs}
\begin{document}

\title{Data-Driven Recursive Least Squares Estimation for Model Predictive Current Control of Permanent Magnet Synchronous Motors 
\thanks{This work was funded by the German Research Foundation (DFG) under the reference number BO 2535/20-1.}
}

\author{

\IEEEauthorblockN{Anian Brosch, 
									S{\"o}ren Hanke, 
									Oliver Wallscheid, 
									Joachim B{\"o}cker }

\IEEEauthorblockA{Power Electronics and Electrical Drives, Paderborn University, 33095 Paderborn, Germany,\\\{brosch, hanke, wallscheid, boecker\}@lea.upb.de}
													
}

\maketitle

\begin{abstract}
The performance of model predictive controllers (MPC)  strongly depends on the model quality. 
In the field of electric drive control, white-box (WB) modeling approaches derived from first-order physical principles are most common. This procedure typically does not cover parasitic effects and parameter deviations are frequent. These issues are particularly crucial in the domain of self-commissioning drives when 
a hand-tailored, accurate WB plant model is not available.
In order to compensate for such modeling errors and, therefore, to improve the control performance during transients and steady-state, this paper proposes a data-driven, real-time capable recursive least squares (RLS) estimation method for the current control of a permanent magnet synchronous motor (\pmsm). 
The effect of the flux linkage and voltage harmonics due to the winding scheme can also be taken into account. Moreover, a compensating scheme for the interlocking time of the inverter is proposed. 
The proposed identification algorithm is investigated using the well-known finite-control-set MPC (FCS-MPC) in the rotor-oriented coordinate system. 
The extensive experimental results show the superior performance of the presented scheme compared to a FCS-MPC-based on a state-of-the-art WB motor model using look-up tables for adressing (cross-)saturation.
\end{abstract}

\section{Introduction } \label{sec:intro}

Due to the increasing computing power, model predictive control (MPC) is gaining popularity in the field of power electronics and electrical drives 
\cite{Cortes.2008, Linder.2010, RodriguezPerez.2012, Vazquez.2014, Kouro.2015}. In the context of field\mbox{-}oriented motor control, the current controller is the important basis of all further control loops. The main requirements for the current controller of an electrical drive is a zero steady state error and - in transient operation - a short settling time with little overshoot. Also secondary requirements such as low current distortion can be formulated.

The performance of MPC depends largely on the accuracy of the model of a given system. 
For current control of a permanent magnet synchronous motor (\pmsm) usually the basic fundamental\mbox{-}wave motor model \cite{RodriguezPerez.2012} with fixed motor parameters is used. If only this model is used, it can lead to large prediction errors and, thus, to a worse performance compared to other control techniques such as linear field\mbox{-}oriented approaches based on proportional\mbox{-}integral (PI) feedback. 
Prediction errors due to model deviations can be caused by several reasons such as varying motor parameters, inverter non\mbox{-}linearities and flux linkage harmonics. 
Motor parameters, e.g. inductances, ohmic resistances, etc. depend on temperature, frequency and (cross\mbox{-})saturation effects \cite{Wallscheid.2014, Peters.2012}. Also production deviations and aging processes can influence the motor parameters \cite{Ott.2016, Huger.2015}. A possible solution to overcome the problem of model deviation due to varying motor parameters is to identify them for all occurring operating conditions and store them in look-up tables (LUT) \cite{CintronRivera.2013}. However, due to the large amount of different root causes it is unrealistic to cover all of them in offline identified LUTs especially motor-specific influences (e.g. production deviations) and those which are changing over time (e.g. material aging). Another approach is to identify the motor parameters online. In the context of sensorless control a broad range of recursive least squares (RLS) methods for estimating a white-box (WB) model with physical parameters and a given structure of electrical machines exist \cite{Ichikawa.2006, Inoue.2009, Omrane.2013}. There, the varying rotor speed and position information is extracted from a WB model which is typically provided by an expert engineer. 

In contrast, the data-driven method presented here does not use a predefined model structure consisting of physical parameters. As a result, the data\mbox{-}driven model allows more flexibility than the WB model and, therefore, can adapt better to the present operation condition. 
The estimator merely identifies the behavior between inputs and outputs in a data\mbox{-}driven fashion. 
A least squares (LS) method to estimate the motor model for a PMSM, also in a data\mbox{-}driven fashion, in the context of model predictive control can be found in \cite{Hanke.2019}. This method estimates one autonomous system for every switching state. After recording a series of measurements online, the LS method can estimate the autonomous systems and update the model to improve the performance of the controller at the present operating point. However, the motor model is not recursively updated after each controller iteration. In comparison to \cite{Hanke.2019}, the motor model presented here is continuously estimated online using a RLS method and an inverter non\mbox{-}linearity compensation is taken into account.

Techniques to compensate non\mbox{-}linearities of inverters are described in \cite{Chen.2015, Choi.1996, Hu.2005, Inoue.2009}, but these require a modulator. 
Since a modulator is not required when using finite\mbox{-}control\mbox{-}set MPC (FCS-MPC), these compensation techniques cannot be used. For this reason, a  compensation scheme for the interlocking time of the inverter in context of FCS-MPC is presented.

The well\mbox{-}known FCS-MPC with a prediction horizon of $N=1$ is a suitable choice for the current control loop because of its low computational complexity and the short settling time with overshoot\mbox{-}free response \cite{Rodriguez.2013}. To compensate for the computation time delay of the FCS-MPC, an additional prediction step at the beginning of each controller iteration is executed \cite{Cortes.2012}. A further advantage of this control method, compared to modulator\mbox{-}based techniques (e.g. continuous\mbox{-}control\mbox{-}set MPC) is the inherent system excitation of the applied switching sequence, which is beneficial for any online identification. Moreover, the proposed identification is independent from the induced voltage in the fundamental\mbox{-}wave domain and, therefore, can be used in the entire speed range including standstill. Compared to any other indirect control approach utilizing a modulator with regular sampling, no additional signal injection is required for the proposed control and identification scheme making it particularly suitable for self\mbox{-}commissioning drive applications.

\section{Mathematical Model}

The following equations summarize the state\mbox{-}of\mbox{-}the\mbox{-}art WB motor model, the effect of the inverter interlocking time and the FCS-MPC.
  \subsection{Coordinate Systems}
	Transformations between stator fixed three\mbox{-}phase \textit{abc} and the stator fixed $\alpha \beta$ coordinate system can be done with the following matrices
	\begin{equation}
	\begin{aligned}
	\Tababc &= \frac{2}{3}\begin{bmatrix} 1 & -\frac{1}{2} & -\frac{1}{2} \\ 0 & \frac{\sqrt{3}}{2} & -\frac{\sqrt{3}}{2} \ \end{bmatrix},\ \ \; \,  \Tabcab=\Tababc^\dagger.
	\label{eq:Transformations_abcab}
	\end{aligned}
	\end{equation}
	Here, $\dagger$ denotes the Moore\mbox{-}Penrose pseudoinverse and bold symbols depict matrices/vectors.
In a similar way the transformations between the $\alpha \beta$ and rotor fixed \textit{dq} coordinate system can be formulated
	\begin{equation}
	\begin{aligned}
	\Tdqab(\eel(t)) &= \begin{bmatrix} \cos(\eel(t)) & \sin(\eel(t)) \\ -\sin(\eel(t)) & \cos(\eel(t)) \ \end{bmatrix}, \\
	\Tabdq(\eel(t)) &=\Tdqab^{-1}(\eel(t)),
	\label{eq:Transformations_dqab}
	\end{aligned}
	\end{equation}
	in which $\eel$ denotes the electrical rotation angle of the PMSM. 
	With \eqref{eq:Transformations_abcab} and \eqref{eq:Transformations_dqab} the transformation between the \textit{abc} and \textit{dq} coordinate system can be obtained
	\begin{equation}
	\begin{aligned}
	\Tdqabc(\eel(t)) &= \Tdqab(\eel(t))\Tababc, \\
	\Tabcdq(\eel(t)) &=\Tdqabc^\dagger(\eel(t)).
	\label{eq:Transformations_dqabc}
	\end{aligned}
	\end{equation}

\subsection{Differential Equations of PMSM} \label{sec:ode_PMSM}
Starting from a general motor model, a simplified motor model is derived to discuss the influence of (cross-)saturation effects and small\mbox{-}angle approximation on the density/sparsity of the state\mbox{-}space system matrices.
\subsubsection{Generalized Fundamental\mbox{-}Wave Model}
The current\mbox{-}based discrete\mbox{-}time difference equation of a PMSM considering  (cross-)saturation effects in the stator\mbox{-}based \textit{dq} coordinate system can be described as follows \cite{Peters.2014}
\begin{equation}
	\begin{aligned}
	\idq[k+1]  =& \left(\bm{I}-\Ldqdiffm^{-1}\Rs\Ts\right) \idq[k] \\
	&+\Ldqdiffm^{-1}\Tabdq(-\Ts\wel)\Ts\udq[k]\\
	&+\Ldqdiffm^{-1}\left(\Tabdq(-\Ts\wel)-\bm{I}\right) \Psidq[k].
	\label{eq:RKV1_idq_PMSM}
	\end{aligned}
	\end{equation}
Above, $\idq=\begin{bmatrix} \id & \iq\end{bmatrix}^\Tr$ is the current, $\udq=\begin{bmatrix} \ud & \uq\end{bmatrix}^\Tr$ the stator voltage, $\Psidq=\begin{bmatrix} \Psid & \Psiq\end{bmatrix}^\Tr$ the flux linkage, $\Rs$ the stator resistance, $\wel$ the electrical angular velocity, $\Ts$ the constant time interval between two samples, $\bm{I}$ the unit matrix and $\Ldqdiffm$ the differential inductance matrix defined by
\begin{equation}
\Ldqdiffm(\idq,\wel)[k]=\begin{bmatrix} \Ldddiff[k] & \Ldqdiff[k] \\ \Lqddiff[k] & \Lqqdiff[k] \end{bmatrix}.
\end{equation}
The differential inductance matrix is a function of the current due to (cross-)saturation effects and the electric angular velocity due to iron losses.
Due to the same effects, the flux is also a nonlinear function of the current and the electric angular velocity $\Psidq=\Psidq(\idq,\wel)$. 
The discrete\mbox{-}time difference equation \eqref{eq:RKV1_idq_PMSM} can be formulated in state\mbox{-}space representation
	\begin{equation}
	\begin{aligned}
	\idq&[k+1] = \A \idq[k] + \B \udq[k] + \E \\
	\text{with: } \A =& \left(\bm{I}-\Ldqdiffm^{-1}\Rs\Ts\right)=\begin{bmatrix} a_{\rm 11} & a_{\rm 12} \\ a_{\rm 21} & a_{\rm 22} \end{bmatrix}, \\
								a_{\rm 11}=& 1-\frac{\Lqqdiff}{|\Ldqdiffm|}\Rs\Ts,\ \ \,
								a_{\rm 12}= \frac{\Ldqdiff}{|\Ldqdiffm|}\Rs\Ts,\\
								a_{\rm 21}=& \frac{\Lqddiff}{|\Ldqdiffm|}\Rs\Ts,\ \ \,
								a_{\rm 22}= 1-\frac{\Ldddiff}{|\Ldqdiffm|}\Rs\Ts,\\
	              \B =& \Ldqdiffm^{-1}\Tabdq(-\Ts\wel)\Ts=\begin{bmatrix} b_{\rm 11} & b_{\rm 12} \\ b_{\rm 21} & b_{\rm 22}\end{bmatrix}, \\
								b_{\rm 11}=& \frac{\Lqqdiff\cos(-\Ts\wel)-\Ldqdiff\sin(-\Ts\wel)}{|\Ldqdiffm|}\Ts,\\
								b_{\rm 12}=& -\frac{\Ldqdiff\cos(-\Ts\wel)+\Lqqdiff\sin(-\Ts\wel)}{|\Ldqdiffm|}\Ts,\\
								b_{\rm 21}=& -\frac{\Lqddiff\cos(-\Ts\wel)-\Ldddiff\sin(-\Ts\wel)}{|\Ldqdiffm|}\Ts,\\
								b_{\rm 22}=& \frac{\Ldddiff\cos(-\Ts\wel)+\Lqddiff\sin(-\Ts\wel)}{|\Ldqdiffm|}\Ts,\\
							  \E =& \Ldqdiffm^{-1}\left(\Tabdq(-\Ts\wel)-\rm{\bm{I}}\right) \Psidq[k]=\begin{bmatrix} e_{\rm 1} \\ e_{\rm 2} \end{bmatrix}, \\
								e_{\rm 1}=& -\frac{(\Ldqdiff\sin(-\Ts\wel)-\Lqqdiff(\cos(-\Ts\wel)-1))\Psid}{|\Ldqdiffm|}\\
								          & -\frac{(\Lqqdiff\sin(-\Ts\wel)+\Ldqdiff(\cos(-\Ts\wel)-1))\Psiq}{|\Ldqdiffm|},\\
								e_{\rm 2}=& \frac{(\Ldddiff\sin(-\Ts\wel)-\Lqddiff(\cos(-\Ts\wel)-1))\Psid}{|\Ldqdiffm|}\\
								          & \frac{(\Lqddiff\sin(-\Ts\wel)+\Ldddiff(\cos(-\Ts\wel)-1))\Psiq}{|\Ldqdiffm|}.\\
	\label{eq:discrete_ODE_PMSM}
	\end{aligned}
	\end{equation}
Above, it can be seen that the system matrices $\B$ and $\E$ are in general dense.

\subsubsection{Simplified Fundamental\mbox{-}Wave Model}
If \mbox{(cross-)saturation} effects and iron losses are neglected the flux $\Psidq$ can be expressed with the absolute inductance matrix $\Ldq$ and the permanent magnet flux $\Psip$
\begin{equation}
\Psidq=\Ldq\idq+\Psipv=\begin{bmatrix} \Ld & 0 \\ 0 & \Lq \end{bmatrix}\idq+\begin{bmatrix}    \Psip \\ 0 \end{bmatrix}.
\label{eq:AbsInd}
\end{equation}
In addition to that, the differential inductance matrix is identical with the absolute inductance matrix, such that ${\Ldqdiffm=\Ldq}$.
If a small\mbox{-}angle approximation (small change of angle $\Delta \eel=\Ts \wel$ per sample time $\Ts$, such that $\cos(\Delta \eel)\approx 1$ and $\sin(\Delta \eel)\approx \Delta \eel$) is also applied, the  generalized fundamental\mbox{-}wave model \eqref{eq:discrete_ODE_PMSM} simplifies to the often used discrete state\mbox{-}space representation of the PMSM with the system matrices
	\renewcommand{\arraystretch}{1.3}
	\begin{equation}
	\begin{split}
	\A  &=\begin{bmatrix} 1-\frac{\Rs\Ts}{\Ld} & \frac{\Lq\wel\Ts}{\Ld}\\	-\frac{\Ld\wel\Ts}{\Lq} & 1-\frac{\Rs\Ts}{\Lq} \end{bmatrix}, \\
	\B  &=\begin{bmatrix} \frac{\Ts}{\Ld} & 0 \\ 0 & \frac{\Ts}{\Lq}\end{bmatrix},\ \ \,
	\E =\begin{bmatrix} 0\\ -\frac{\Psip\wel\Ts}{\Lq}\end{bmatrix},
	\label{eq:discrete_ODE_PMSM_simp_1}
	\end{split}
	\end{equation}
	\renewcommand{\arraystretch}{1.0}	
In comparison to the generalized fundamental\mbox{-}wave model \eqref{eq:discrete_ODE_PMSM} the system matrices $\B$ and $\E$ of the simplified fundamental\mbox{-}wave model \eqref{eq:discrete_ODE_PMSM_simp_1} are sparse. 

\subsection{Inverter}
\begin{figure}
    \centering
    \includegraphics[width=0.49\textwidth]{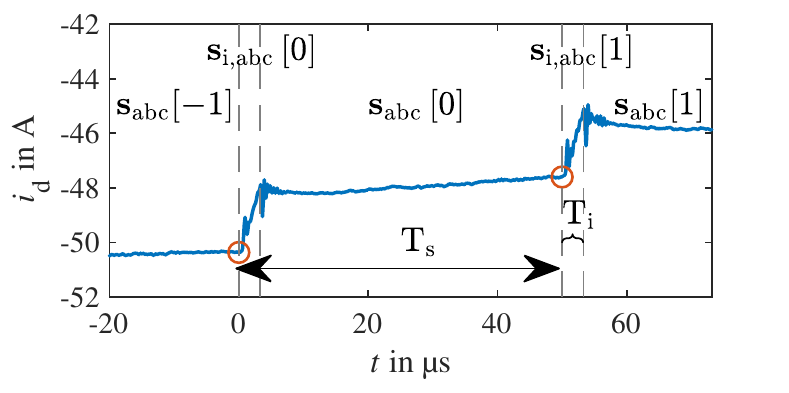}
    \caption{Exemplary trajectory of the $\id$ current, where the effect of the intermediate vectors $\siabc[0]$ and $\siabc[1]$ during the interlocking time $\Ti$ can be seen. Also the sample points (red) and the sample time $\Ts$ of the controller is depicted.}
    \label{fig:Intermediate_Vector_Example}
\end{figure}

\begin{figure}
    \centering
    \includegraphics[width=0.09\textwidth]{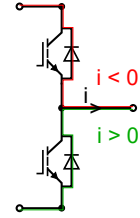}
    \caption{In this definition a positive current remains to an active lower with an inactive upper switch and negative current vice versa during the interlocking time.}
    \label{fig:Inverter_Phase_Lag}
\end{figure}

The finite set of the voltages in the $\alpha \beta$ coordinate system due to the finite number of possible switching combinations of the three-phase two\mbox{-}level inverter can be defined as follows
	\begin{equation}
	\uab \in \left\{ \uab \in \mathbb{R}^2 \vert \uab = \UDC\Tababc \sabc \right\} 
	\label{eq:Set_U_ab}
	\end{equation}
	where $\UDC$ is the DC\mbox{-}link voltage and $\sabc$ is the switching state of the inverter defined by 
	\begin{equation}
	\sabc =\begin{bmatrix} s_\mathrm{a} & s_\mathrm{b} & s_\mathrm{c} \end{bmatrix}^\Tr \in \left\{0,1\right\}^3 \subset  	\mathbb{Z}^3.
	\label{eq:sn}
	\end{equation}
	With the equations above, the stator voltage $\udq$ can be calculated
  \begin{equation}
	\udq= \Tdqab(\eel)\uab=\UDC\Tdqabc(\eel)\sabc.
	\label{eq:udquab}
	\end{equation}	
	To avoid a hard short circuit in the inverter the interlocking time $\Ti$ is passed until the transistor is turned on if the other transistor in a phase leg was previously in conducting mode. During this time, both transistors are off and the phase current flows through one of the two diodes. Depending on the sign of the current, the phase of the PMSM is then connected to the upper or lower DC-link potential via the conducting diode (see Fig. \ref{fig:Inverter_Phase_Lag}). In the following, the resulting switching state shall be denoted as intermediate switching vector $\siabc$  (see Fig. \ref{fig:Intermediate_Vector_Example}). 
The number of zero crossings of a phase current during the interlocking time is assumed to be negligible small. 
The \ivector $\siabc$ at the $k$th time instant without discontinuous conduction can be computed as follows
	\begin{equation}
	\begin{aligned}
	\siabc[k] =& \left| \sabc[k]-\sabc[k-1] \right|\odot h \left( -\iabc[k]\right) \\
	          &+\left|\left| \sabc[k]-\sabc[k-1] \right|-\boldsymbol{1}\right|\odot \sabc[k]
	\label{eq:Compute_ivector}
	\end{aligned}
	\end{equation}
with the heaviside function $h$, the elementwise multiplication $\odot$ and $\boldsymbol{1}=\begin{bmatrix} 1 & 1 & 1 \end{bmatrix}^\Tr$. 
Usually the effect of the interlocking time is neglected for FCS-MPC, which leads to a systematic modeling and as a consequence control error.
	
\subsection{Finite-Control-Set MPC}
After the calculation of the intermediate vector has been shown, it can be integrated into the prediction and thus into the optimization problem of the FCS\mbox{-}MPC. The cost function for the FCS\mbox{-}MPC with prediction horizon of $N=1$ is designed to penalize the squared Euclidean distance between the predicted current of the next time step and the reference current $\idqref$.
With the definition of the cost function the optimization problem can be formulated
\begin{subequations}
\label{eq:opt_controller}
	\begin{align}
			&	\min_{s_{\mathrm{abc}}[k]} & & \norm{\idq[k+1]-\idqref}^2_2
			&                              & & \notag \\ 
			&	\mathrm{\ s.t.}              & & \udq[k]=\UDC\Tdqabc(\eel[k])\sabc[k], \\
			&                              & & \uidq[k]=\UDC\Tdqabc(\eel[k])\siabc[k], \\
			&                              & & \udqm[l\vert k]=\frac{\Ts-\Ti}{\Ts}\udq[k]+\frac{\Ti}{\Ts}\uidq[k], \label{eq: averageU}\\
			&														   & & \idq[\vert k+1] = \A \idq[k] + \B \udqm[k] + \E.
	\end{align}			
\end{subequations}
In \eqref{eq: averageU} the voltages of the intermediate and the switching vectors are averaged.
Thanks to the finite set of the switching states the cost function can be evaluated for all feasible switching states $\sabc^T[k]$. The switching state, from which a minimum cost results, is denoted as $\sabc^*[k]$. This switching state is applied and the procedure must be repeated at the next sampling instant. 

\section{Experimental Test Setup}

The FCS\mbox{-}MPC in combination with the system identification have been implemented on a laboratory test bench. The electrical drive system is a highly utilized interior permament magnet synchronous motor (Brusa: HSM1\mbox{-}6.17.12\mbox{-}C01) for automotive applications and a 2\mbox{-}level IGBT inverter (Semikron: 3×SKiiP 1242GB120\mbox{-}4D). Additionally, a speed\mbox{-}controlled induction machine as load motor (Schorch: LU8250M\mbox{-}AZ83Z\mbox{-}Z) is mechanically coupled with the test motor. The testbench is further equipped with a dSPACE DS1006MC rapid\mbox{-}control\mbox{-}prototyping system.
The most important test motor, inverter and control parameters are listed in Tab. \ref{tab:parm}.

\section{Offline Ordinary Least Squares} \label{sec:OLS}

Before various models for online system identification are tested in a closed control loop, a data set  was recorded for offline pre\mbox{-}investigations. This data set is available as supplementary material \cite{Data_Set_RLS.}. The FCS\mbox{-}MPC \eqref{eq:opt_controller} was used as control algorithm for recording the data set, whereby \eqref{eq:RKV1_idq_PMSM} was applied as prediction model. Here, the entries of the differential inductance matrix $\Ldqdiffm$ and the flux vector $\Psidq$ were continuously updated with the use of LUTs in dependence on $\idq$. The data set was recorded at the rated current operating point $\idqref=\begin{bmatrix} -170 & 170 \end{bmatrix}^\Tr\SI{}{A}$ at $\SI{2000}{min^{-1}}$ and contains $40000$ samples which corresponds to a measuring time of $\SI{2}{s}$. The quality of different models was investigated offline on the basis of the data set using the ordinary least squares (OLS) method. Here the quadratic deviation between measurement and model 
\begin{equation}
	\bm{r} = {\bm{\Psi}}-\bm{\Xi}\bm{\theta}.
	\label{eq:comp_resid}
\end{equation}
is minimized. Above, $\bm{\Psi}$ denotes to the measurement vector, $\bm{\Xi}$ to the regressor matrix and $\bm{\theta}$ to the parameter vector. As cost function the squared Euclidean norm of the residuals $ {J(\bm{\theta})=\norm{\bm{r}}^2_2=\left(\bm{\Psi}-\bm{\Xi}\bm{\theta}\right)^\Tr\left(\bm{\Psi}-\bm{\Xi}\bm{\theta}\right)}$ is used. The optimal parameter vector ${\bm{\theta}^*}$ can than be calculated with the following equation \cite{Walter.1997}
\begin{equation}\label{eq:LS_siso_stat_001}
	\bm{\theta}^* = \left(\bm{\Xi}^\Tr\bm{\Xi}\right)^{-1}\bm{\Xi}^\Tr\bm{\Psi} = \bm{\Xi}^\dagger\bm{\Psi}.
\end{equation}
To identify the entries of the system matrices $\A$, $\B$, and $\E$, state\mbox{-}space representation \eqref{eq:discrete_ODE_PMSM} can be rewritten into two  separate least squares problems.
\begin{align}
&\underbrace{\begin{bmatrix}\id[2] \\ \vdots \\ \id[M+1]\end{bmatrix}}_{\boldsymbol{\Psi}_{\rm{d}}} = \notag \\  &\underbrace{\begin{bmatrix} \id[1] & \iq[1] & \udo[1] & \uqo[1] & 1  \\ \vdots & \vdots & \vdots &  \vdots & \vdots \\ \id[M] & \iq[M] & \udo[M] & \uqo[M] & 1\end{bmatrix}}_{\bm{\Xi}_{\rm{d}}} \underbrace{\begin{bmatrix} a_{\rm 11} \\ a_{\rm 12} \\ b_{\rm 11} \\ b_{\rm 12} \\ e_{\rm 1} \end{bmatrix}}_{\bm{\theta}_d} + \underbrace{\begin{bmatrix}r_{\rm{d}}[1] \\ \vdots \\ r_{\rm{d}}[N]\end{bmatrix}}_{\rdv} \label{eq: OLSd}
\end{align}
and
\begin{align}
&\underbrace{\begin{bmatrix}\iq[2] \\ \vdots \\ \iq[M+1]\end{bmatrix}}_{\boldsymbol{\Psi}_{\rm{q}}} = \bm{\Xi}_q \underbrace{\begin{bmatrix} a_{\rm 21} \\ a_{\rm 22} \\ b_{\rm 21} \\ b_{\rm 22} \\ e_{\rm 2} \end{bmatrix}}_{\bm{\theta}_q} + \underbrace{\begin{bmatrix}r_{\rm{q}}[1] \\ \vdots \\ r_{\rm{q}}[N]\end{bmatrix}}_{\rqv}. \label{eq: OLSq}
\end{align}
The constant $e_{\rm{1}}$ can be estimated with the help of the pseudoregressor $1$ in the regression matrix.
After calculating the parameter vectors ${\bm{\theta}^*_{\rm{d}}}$ and ${\bm{\theta}^*_{\rm{q}}}$ 
the residuals $\rdv$, $\rqv$ can than be computed with \eqref{eq:comp_resid}.

\subsection{Residual Analysis} \label{sec:ResidAnalysis}
\begin{figure} 
    \centering
    \includegraphics[width=0.49\textwidth]{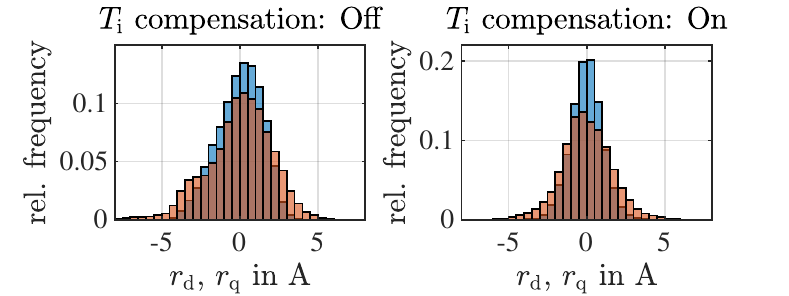}
    \caption{Histogram of the residuals in $\rdv$ (orange) and $\rqv$ (blue) without (left) and with (right) interlocking time compensation.}
    \label{fig:Regressor_Analysis_hist}
\end{figure}

\begin{figure} 
    \centering
    \includegraphics[width=0.49\textwidth]{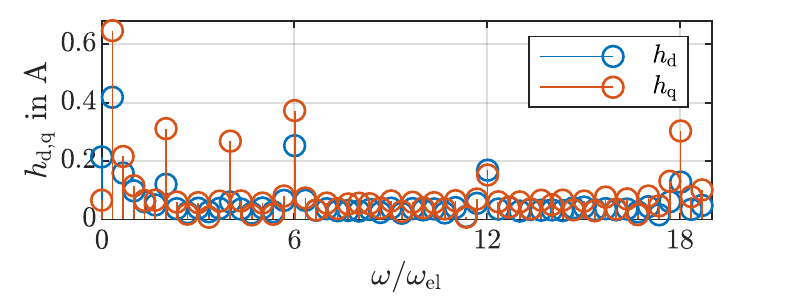}
    \caption{Fourier coefficients $h_{\rm d}$ and $h_{\rm q}$ of the residuals $\rdv$ and $\rqv$ over electrical orders with active interlocking time compensation.}
    \label{fig:Regressor_Analysis_FFT}
\end{figure}

Residual analysis plays an important role in validating the regression model. There are many statistical tools for model validation, but the primary tool for most modeling applications is graphical residual analysis \cite{NIST.}. Often zero mean and normally distributed  noise is assumed for the residuals. The histogram plot of the residuals can be used to check if the residuals are normally distributed. In Fig. \ref{fig:Regressor_Analysis_hist} the residuals with and without interlocking time compensation can be seen. With interlocking time compensation the variance can be reduced and also the symmetry of the distribution is improved. In electrical drives there are also harmonics due to the motor geometry. For example, non\mbox{-}ideal winding schemes or asymmetries result in a nonideal sinusoidal flux distribution in the air gap. The fundamental wave models \eqref{eq: OLSd} and \eqref{eq: OLSq} are not able to predict these current harmonics. For this reason a discrete fourier transform of the residuals is shown in Fig. \ref{fig:Regressor_Analysis_FFT}. Here, the characteristic 6\textit{th}, 12\textit{th} and 18\textit{th} harmonics in the \textit{dq} coordinate system due to the winding scheme as well as the 3\textit{th} subharmonic can be seen. Due to the pole pair number of 3, this subharmonic corresponds to the fundamental frequency of the mechanical system. One reason for this subharmonic can be asymmetries between the pole pair systems. 
\begin{table}[ht] 
				\caption{PMSM, inverter and control parameters}
			 \label{tab:parm}
				\centering
				\begin{tabular}{l|c|c}
				\hline
				Nominal power                    		& $P_{\rm{mech}}$   		     			  	& \SI{55}{kW}\\
				Pole pair number                    & $p$               & 3                         \\
				DC-link voltage                  		& $u_\mathrm{DC}$   							  	& \SI{300}{V}\\
				Nominal current                   	& $I$             								  	& \SI{170}{A}\\ 
				\hline
				Inverter interlocking time          & $\Ti$   								            & \SI{3.3}{\us} \\
				Controller cycle time    	          & $\Ts$   								            & \SI{50}{\us} \\
				MPC prediction horizon    		      & $N$   	  					             		& 1\\											
				\hline
				\end{tabular}
		\end{table}	

\section{Regression Models}

In the following different model structures are defined to compare the performance in the offline OLS method and later also in the online RLS method.

\subsubsection{Dense Fundamental\mbox{-}Wave Model}
The dense fundamental\mbox{-}wave (DFW) model is able to map the behavior of the generalized motor model \eqref{eq:RKV1_idq_PMSM} with dense system matrices and (cross-)saturation effects.
The $m$th line of the regressor matrix $\bm{\Xi}_{\rm{d,DFW}}$ and $\bm{\Xi}_{\rm{q,DFW}}$ are defined as in \eqref{eq: OLSd} and \eqref{eq: OLSq}
\begin{align}
\bm{\xi}_{\rm{d,DFW}}[m]^\Tr &= \begin{bmatrix} \id[m] & \iq[m] & \ud[m] & \uq[m] & 1 \end{bmatrix} \label{eq: OLSd_BB_PV}, \\
\bm{\xi}_{\rm{q,DFW}}[m]^\Tr &=\bm{\xi}_{\rm{d,DFW}}[m]^\Tr \label{eq: OLSq_BB_PV}.
\end{align}
The parameter vectors are
\begin{align}
\bm{\theta}_{\rm{d,DFW}} =& \begin{bmatrix} a_{\rm 11} & a_{\rm 12} & b_{\rm 11} & b_{\rm 12} & e_{\rm 1} \end{bmatrix}^\Tr,  \\
\bm{\theta}_{\rm{d,DFW}} =& \begin{bmatrix} a_{\rm 21} & a_{\rm 22} & b_{\rm 21} & b_{\rm 22} & e_{\rm 2} \end{bmatrix}^\Tr  .
\end{align}
For reasons of simplicity, the index $m$ will not be used for the future definition of the models.

\subsubsection{Sparse Fundamental\mbox{-}Wave Model}
In comparison to the DFW model the parameters $b_{\rm 12}$, $b_{\rm 21}$ and $e_{\rm 1}$ are set to zero for the sparse fundamental\mbox{-}wave (SFW) model. Thus, this model structure can only represent a model with sparse system matrices $\B$ and $\E$. For this reason, no cross-saturation effects can be modeled, see Sec. \ref{sec:ode_PMSM}. 
The $m$th line of the regressor matrix $\bm{\Xi}_{\rm{d,SFW}}$ and $\bm{\Xi}_{\rm{q,SFW}}$ are defined as 
\begin{equation}
\bm{\xi}_{\rm{d,SFW}}^\Tr = \begin{bmatrix} \id & \iq & \ud \end{bmatrix}, \ \ \,  \bm{\xi}_{\rm{q,SFW}}^\Tr = \begin{bmatrix} \id & \iq & \uq & 1 \end{bmatrix} \label{eq: OLSq_GB_RM}.
\end{equation}
The parameter vectors are
\begin{align}
\bm{\theta}_{\rm{d,SFW}} =& \begin{bmatrix} a_{\rm 11} & a_{\rm 12} & b_{\rm 11}  \end{bmatrix}^\Tr \label{eq: OLSd_GB_PV}, \\
\bm{\theta}_{\rm{q,SFW}} =& \begin{bmatrix} a_{\rm 21} & a_{\rm 22} & b_{\rm 22} & e_{\rm 2} \end{bmatrix}^\Tr \label{eq: OLSq_GB_PV}.
\end{align}

\subsubsection{Augmented Fundamental\mbox{-}Wave Model}
As seen in section \ref{sec:ResidAnalysis}, the residuals still contain deterministic components in the form of harmonics. In order to identify a model that can represent harmonics, sine and cosine terms must be added, in a similar way as in \cite{Wang.2014}, to the regressor matrix with the respective frequency of the harmonics. The number of parameters $n$ to be estimated increase with each additional harmonic. 
For reasons of the real-time capability of the online system identification method, it is important to find a trade-off between a limited number of parameters and model accuracy. A suitable method for parameter shrinkage and selection is LASSO regression \cite{Tibshirani.1996}.
For the LASSO method regressors of the DFW model \eqref{eq: OLSd_BB_PV} and \eqref{eq: OLSq_BB_PV}, as well as these multiplied by sine and cosine terms of order $z$, were used
\begin{align}
\bm{\xi}_{\rm{d,LASSO}}^\Tr =& \begin{bmatrix} \bm{\xi}_{\rm{d,DFW}}^\Tr & \sin(z\eel)\bm{\xi}_{\rm{d,DFW}}^\Tr & \cos(z\eel)\bm{\xi}_{\rm{d,DFW}}^\Tr \end{bmatrix},  \\
\bm{\xi}_{\rm{q,LASSO}}^\Tr =& \begin{bmatrix} \bm{\xi}_{\rm{q,DFW}}^\Tr & \sin(z\eel)\bm{\xi}_{\rm{q,DFW}}^\Tr & \cos(z\eel)\bm{\xi}_{\rm{q,DFW}}^\Tr \end{bmatrix} .
\end{align}
The most important regressors can now be selected using LASSO regression. This was repeated for different dominant orders, see Fig. \ref{fig:Regressor_Analysis_FFT}. After selection of the most important regressors the $m$th line of the regressor matrix $\bm{\Xi}_{\rm{d,AFW}}$ and $\bm{\Xi}_{\rm{q,AFW}}$ can be defined as
\renewcommand{\arraystretch}{1.1}
\begin{align}
\bm{\xi}_{\rm{d,AFW}}= \begin{bmatrix} \id \\ \iq \\ \ud \\ \ud\sin\left(\frac{1}{3}\eel\right) \\ \ud\cos\left(\frac{1}{3}\eel\right) \\ \ud\sin\left(6\eel\right) \\ \ud\cos\left(6\eel\right)\end{bmatrix},\ \ \, 
\bm{\xi}_{\rm{q,AFW}} = \begin{bmatrix} \id  \\ \iq  \\ \uq  \\ 1 \\ \uq\sin\left(\frac{1}{3}\eel \right) \\ \uq\cos\left(\frac{1}{3}\eel \right) \\ \uq\sin\left(6\eel \right) \\ \uq\cos\left(6\eel \right)\end{bmatrix} \label{eq: OLSq_AGB_RM}.
\end{align}
\renewcommand{\arraystretch}{1.0}
The parameter vectors are
\begin{align}
\bm{\theta}_{\rm{d,AFW}} =& \begin{bsmallmatrix} a_{\rm 11} & a_{\rm 12} & \tilde{b}_{\rm 11,0} & \tilde{b}_{\rm 11,1}  & \tilde{b}_{\rm 11,2} & \tilde{b}_{\rm 11,3} & \tilde{b}_{\rm 11,4}  \end{bsmallmatrix}^\Tr, \label{eq: OLSd_AGB_PV} \\
\bm{\theta}_{\rm{q,AFW}} =& \begin{bsmallmatrix} a_{\rm 21} & a_{\rm 22} & \tilde{b}_{\rm 22,0} & e_{\rm 2} & \tilde{b}_{\rm 22,1}& \tilde{b}_{\rm 22,2}& \tilde{b}_{\rm 22,3}& \tilde{b}_{\rm 22,4}\end{bsmallmatrix}^\Tr \label{eq: OLSq_AGB_PV}.
\end{align}
The matrix elements of the system matrix $\B$ are now time\mbox{-}variant. 
It is interesting to note that although a highly utilized drive with significant saturation and cross-saturation effects is used, the parameters $b_{\rm 12}$, $b_{\rm 21}$ and $e_{\rm 1}$ are set to zero for this operating point with LASSO regression. This is an indication that in this operating point modelling of cross-saturation has a smaller influence on prediction quality than modelling of harmonics. A big disadvantage of this model, however, is that there is not enough excitation to identify the parameters for the sine and cosine regressors for low speeds. If the speed tends to zero, the regressor matrix becomes rank-deficient and, therefore, the parameters cannot uniquely determined.

\subsection{Performance Comparison Utilizing an Offline-OLS}
In order to compare the quality of the different models and the effect of the interlocking time compensation, the residuals for the \textit{d}\mbox{-} and \textit{q}\mbox{-}axis with and without interlocking time compensation were calculated using the data set of Sec. \ref{sec:OLS} \cite{Data_Set_RLS.}.  
Performance criteria are the mean $\mu({r_{\rm{d,q}}})$ and the standard deviation $\sigma(r_{\rm{d,q}})$ of the residuals. Additionally the coefficient of determination $R^2_{\rm d,q}$ is evaluated. The values for these criteria are included in Tab. \ref{tab:performance_comparison_offline_OLS}.
It can be seen that for every model the interlocking time compensation can improve the values for every performance criteria. Even though the mean values of the residuals in the \textit{d}-axis for the SWF and AFW models are higher than that of the DFW model, these deviations are in the range of a few milliampere and, therefore can be considered as negligible. In consequence it can be said that the residuals of all models are bias-free.
The coefficient of determination for every model with interlocking time compensation is larger than $0.97$. This indicates a pretty good model quality. Especially $R^2_{\rm d}=0.993$ for the DFW model with interlocking time compensation in the \textit{d}\mbox{-}axis  is excellent.
\renewcommand{\arraystretch}{1.4}
\begin{table}[ht]
    \centering 	   
    \caption{OLS Performance Comparison at Rated Motor Operation}
    \begin{tabular}{l | c c | c c | c c }
        \hline
        \textbf{Model} & \multicolumn{2}{c|}{DFW} &  \multicolumn{2}{c|}{SFW} &  \multicolumn{2}{c}{ASFW} \\
        $\Ti$ comp. & Off & On &  Off & On & Off & On \\
				\hline
				\multicolumn{7}{c}{\textbf{\textit{d}-Axis}} \\
				\hline\hline
				$\mu({\resd})$ in mA & 0.4 & 0.0 & 20.2 & 9.9 & 20.8 & 10.1  \\
				$\sigma(\resd)$ in A        & 1.443 & 1.021 & 1.572 & 1.064 & 1.414 & 0.937  \\
				$R^2_{\rm d}$            & 0.986 & 0.993 & 0.940 & 0.971 & 0.941 & 0.972  \\
				\hline
				\multicolumn{7}{c}{\textbf{\textit{q}-Axis}} \\
				\hline
				\hline
				$\mu({\resq})$ in mA & 0.0 & -0.3 & -0.2 & 0.0 & 0.0 & 0.0  \\
				$\sigma(\resq)$ in A        & 2.043 & 1.598 & 2.053 & 1.601 & 1.865 & 1.364  \\
				$R^2_{\rm q}$            & 0.971 & 0.982 & 0.970 & 0.982 & 0.976 & 0.987 \\
        \hline
    \end{tabular}
    \label{tab:performance_comparison_offline_OLS}
\end{table}
\renewcommand{\arraystretch}{1.0}

\section{Recursive Least Squares}

The OLS method was used to compare different models on the basis of a data set offline. In contrast to the OLS method used offline, the RLS method is now used for online system parameter identification. The model parameters are recursively corrected with each new measurement. In addition, past measurements can be weighted weaker so that present measurements are taken more into account for parameter estimation.
The calculations that must be executed in each iteration can be formulated as follows \cite{Isermann.2011}
\begin{align}
\bm{\gamma}[k] & = \frac{\bm{P}[k]\bm{\xi}[k+1]}{\lambda+\bm{\xi}^T[k+1]\bm{P}[k]\bm{\xi}[k+1]} \label{eq: correct_term}\\
\hat{\bm{\theta}}[k+1] &= \hat{\bm{\theta}}[k] + \bm{\gamma}[k]\left(\psi[k+1] - \bm{\xi}[k+1]^T\hat{\bm{\theta}}[k]\right)\\
\bm{P}[k+1] &= \frac{1}{\lambda}\left(\bm{I} - \bm{\gamma}[k]\bm{\xi}^T[k+1]\right)\bm{P}[k] .
\end{align}
Above, $\bm{P}$ is a matrix proportional to the covariance matrix $\rm{Cov}(\hat{\bm{\theta}},\hat{\bm{\theta}})$ of the estimated parameter vector and $\lambda$  is a weighting factor that gives a weaker weighting to the past measurements $\bm{\xi}$ of the regressors. In \eqref{eq: correct_term} a correction term is calculated with which the parameter vector $\hat{\bm{\theta}}$ and matrix $\bm{P}$ are updated. As tuning parameter the weighting factor $0<\lambda<1$ can be used. An initial parameter vector $\hat{\bm{\theta}}[0]$ and an initial matrix $\bm{P}[0]$ must also be specified.

\section{Experimental Results}

In order to test the accuracy of the system identification using the RLS method, the behavior in stationary and transient operation was evaluated. The system matrices identified online were used in the closed control loop to compensate the computation time delay and also as prediction model for the FCS\mbox{-}MPC algorithm.
For the RLS method the initial parameter vectors $\hat{\bm{\theta}}_{\rm{d,q}}[0]$ can be computed either with data sheet parameters (if available) or nameplate information in addition with simplifying assumptions. The forgetting factors were set to $\lambda_{\rm{d,q}}=0.99$ for all models. The matrices $\bm{P}_{\rm{d,q}}[0]$ were selected as unit matrices. The term residuals refers from now on to the one-step prediction error using the data\mbox{-}driven RLS models.

\subsection{Steady-State Behavior}
\begin{figure}
    \centering
    \includegraphics[width=0.49\textwidth]{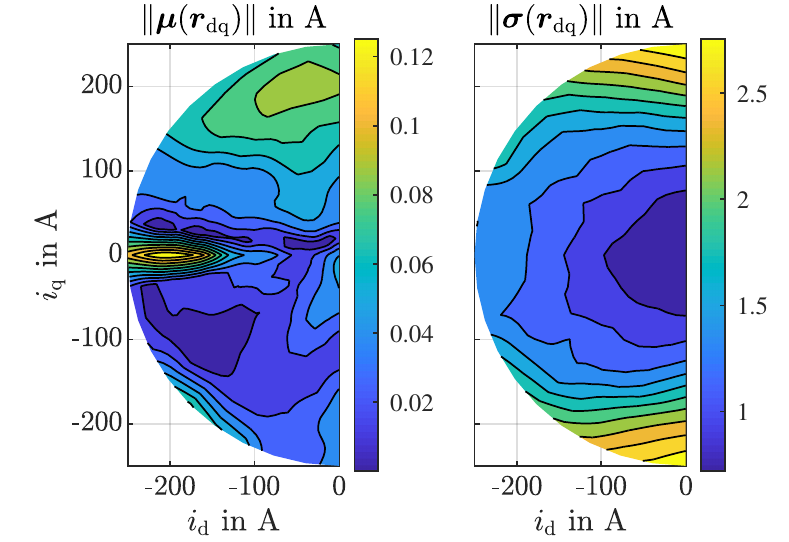}
    \caption{Length of the mean $\norm{\boldsymbol{\mu}}$ and standard deviation vector $\norm{\boldsymbol{\sigma}}$ of the residuals $\resd$, $\resq$ in the left $\id - \iq$ half plane for the DFW model with interlocking time compensation.}
    \label{fig:RLS_std}
\end{figure}

\begin{figure}
    \centering
    \includegraphics[width=0.49\textwidth]{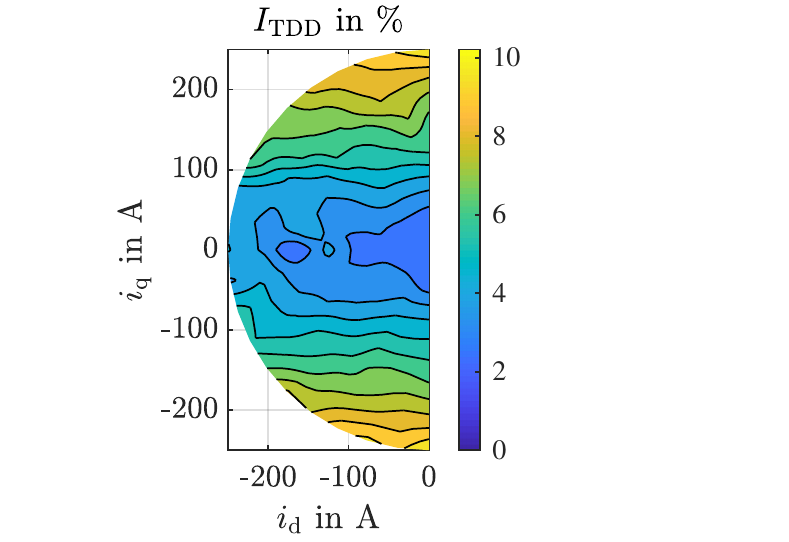}
    \caption{TDD of the phase currents in the left $\id - \iq$ half plane for the DFW model with interlocking time compensation.}
    \label{fig:RLS_TDD}
\end{figure}

In order to gain a comprehensive impression of the quality of the models in stationary operation, 83 equispaced operating points were recorded in the left $\id$\mbox{-}$\iq$ half\mbox{-}plane with a maximal length $\norm{\idq}$ of $\SI{250}{A}$. Every data set of one operating point contains $40000$ samples which corresponds to a measuring time of $2 \;\SI{}{s}$ at a constant speed of $\SI{2000}{min^{-1}}$. The regression models with and without interlocking time compensation and also a LUT\mbox{-}based FCS\mbox{-}MPC as in Sec. \ref{sec:OLS} are compared for different performance criteria. The LUT\mbox{-}based model is called WB model.
In the following, performance criteria are defined, which are evaluated for each operating point.
The length of the mean vector of the residuals is defined as
\begin{equation}\label{eq:meanrdq}
	 \norm{\boldsymbol{\mu}(\resdq)}  = \sqrt{(\mu(\resd))^2+(\mu(\resq))^2}
\end{equation}
to obtain a scalar quantity which gives an indication of the mean of the residuals.
To measure the standard deviation of the residuals a vector is defined that contains the standard deviation of both residuals $\boldsymbol{\sigma}(\resdq)=\begin{bmatrix} \sigma(\resd) & \sigma(\resq) \end{bmatrix}^\Tr$. The length of the vector $\norm{\boldsymbol{\sigma}}$ is used to obtain a scalar value here as well.
To evaluate the steady-state accuracy of the current controller in dependence of the models, the length of the steady-state control deviation vector is defined
\begin{equation}\label{eq:meanedq}
	 \norm{ \edq}  = \norm{\boldsymbol{\mu}(\idq)-\idqref}.
\end{equation}
A further requirement in steady-state operation is to decrease the harmonic current distortion in order to reduce ohmic as well as thermal losses. A suitable measure for the harmonic current distortion is the total demand distortion (TDD) 
\begin{equation}\label{eq:iTDD}
	I_{\rm{TDD}} = \frac{\sqrt{\sum_{h \neq 1}I^2_{h}}}{I_{\rm{nom}}}
\end{equation} 
in which the nominal current ${I_{\rm{nom}}=\frac{1}{\sqrt{2}}\norm{\idq}=\SI{170}{A}}$ refers to the operating condition at ${\idq=\begin{bmatrix} -170 & 170 \end{bmatrix}^\Tr\SI{}{A}}$. The harmonic components $I_{h}$, can be differentiated into the fundamental component $I_{\rm{1}}$ and  the \textit{h}th harmonic and subharmonic components $I_{h}$ in the \textit{abc} coordinate system. The total harmonic distortion (THD) is referred to the present current rather than the nominal current, for this reason the THD tends to go to infinity for small fundamental components while the TDD remains effectively constant.
For this reason the TDD is a more suitable choice for measuring the current distortion than the THD \cite{Geyer.2012}. In order to obtain scalar values of the performance criteria for the complete left $\id - \iq$ half-plane, these are averaged over all $L=83$ operating points
\begin{equation}\label{eq:meanx}
	\overline{ x } = \frac{1}{L}\sum^{L}_{l=1} x_{l}.
\end{equation}
Tab. \ref{tab:performance_comparison_steady_state} shows the values of the performance criteria. All regression models have a lower mean and standard deviation of the residuals compared to the WB model. Due to the increased prediction accuracy, the regression models achieve an improved control performance compared to the WB model, which results in a lower steady\mbox{-}state control deviation and lower current distortion. Although the mean value of the residuals without interlocking time compensation is lower, this is not a major disadvantage for the models with interlocking time compensation, since they are still in the order of some $100\; \SI{}{mA}$. For the models with interlocking time compensation, there are lower values in the standard deviation and also in the current distortion.
The mean and standard deviation of the residuals for the DFW model with interlocking time compensation can be seen in Fig. \ref{fig:RLS_std} and the current distortion in Fig. \ref{fig:RLS_TDD}.

\renewcommand{\arraystretch}{1.4}
\begin{table*}[t]
    \centering 	   
    \caption{FCS-MPC Performance Comparison for Different Models in Steady-State }
    \begin{tabular}{l | c c | c c | c c | c c}
        \hline
        \textbf{Model} & \multicolumn{2}{c|}{DFW} &  \multicolumn{2}{c|}{SFW} &  \multicolumn{2}{c|}{ASFW} & \multicolumn{2}{c}{WB} \\
        $\Ti$ comp.  & Off & On &  Off & On &  Off & On & Off & On\\
				\hline
				\hline
				$\overline{\norm{\boldsymbol{\mu}(\resdq)}}$ in $\SI{}{A}$        & 0.007 & 0.046 &  0.048 & 0.170 &  0.043 & 0.168 & 2.372 & 1.445\\
				\hline	
				$\overline{\norm{\boldsymbol{\sigma}(\resdq)}}$ in $\SI{}{A}$     & 1.946 & 1.533 &  2.765 & 2.440 &  2.714 & 2.381 & 4.951 & 4.942\\
				\hline																				
			  $\overline{ \norm{ \edq} }$    in $\SI{}{A}$                      & 2.523 & 2.541 &  2.673 & 2.656 &  2.693 & 2.609 & 6.989 & 5.078\\
				\hline																				
				$\overline{\Itdd}$ in \%                                          & 5.75 & 5.71 &  5.83 & 5.79 &  5.89 & 5.82 & 5.99 & 6.02\\
        \hline
    \end{tabular}
    \label{tab:performance_comparison_steady_state}
\end{table*}
\renewcommand{\arraystretch}{1.0}

\subsection{Transient Behavior}
\begin{figure}
    \centering
    \includegraphics[width=0.49\textwidth]{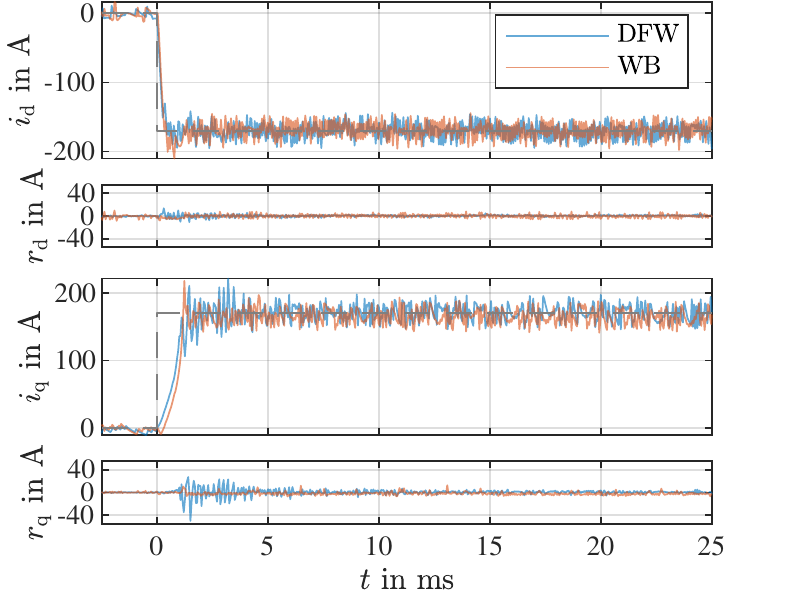}
    \caption{Current step response at constant speed of $2000\;\mathrm{min^{-1}}$. For the DFW as well as the WB model, interlocking time compensation was used.}
    \label{fig:Current_Step_Response}
\end{figure}

The RLS method identifies the system behavior at the present operating point. The identified system matrices are only valid locally and not globally for all operation points. For this reason, the behavior of the RLS method in transient operation with changing operating points is studied. Exemplarily the DFW model with interlocking time compensation is used here. One reason for time\mbox{-}invariant system matrices are changing parameters such as differential and absolute inductances. These inductances have a strong dependency on the current vector $\idq$, especially in the case of highly utilized motors. For this reason, the first scenario to be used is a step response as shown in Fig. \ref{fig:Current_Step_Response}. The short settling time and the overshoot\mbox{-}free behavior of the FCS\mbox{-}MPC can be seen here. During the transient process, the residuals are slightly elevated but decrease towards their steady\mbox{-}state values within a few milliseconds. 

\begin{figure}
    \centering
    \includegraphics[width=0.49\textwidth]{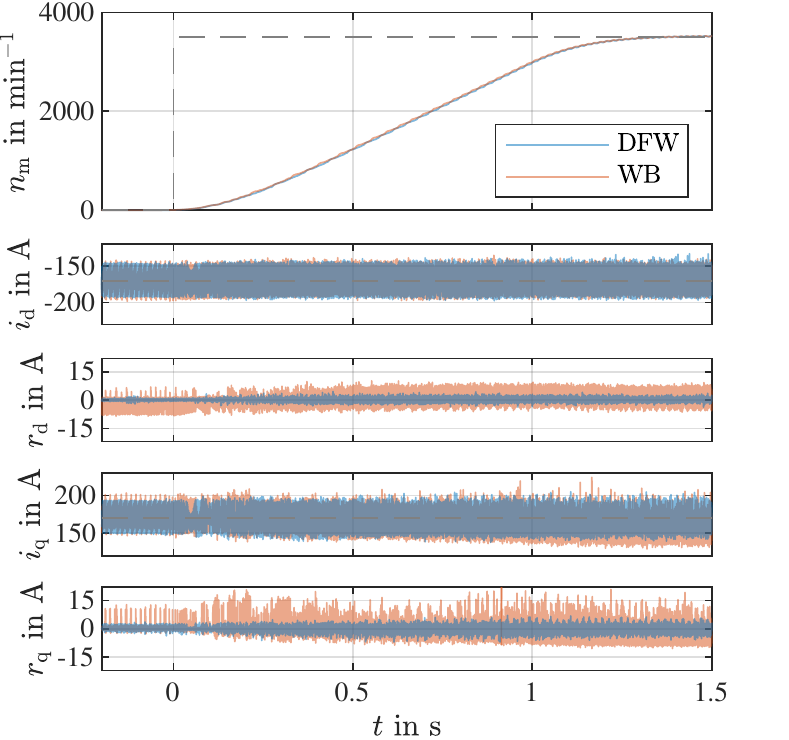}
    \caption{Speed step response at rated current. For the DFW as well as the WB model, interlocking time compensation was used.}
    \label{fig:Velocity_Step_Response}
\end{figure}

Not only the electrical parameters can change, but, also the frequency of the electrical system $\wel$, which appears as a physical parameter in the system matrices. To study the influence of a changing frequency, the behavior of the current controller is investigated during speed transients as shown in Fig. \ref{fig:Velocity_Step_Response}. After reaching the constant cutoff rotational speed, the absolute values of the residuals for the DFW model do not decrease.
This is an indication that no increase in the residuals occurs during a speed change. The reason for this is the large time constant of the mechanical system compared to the fast adaptation of the RLS method in the electrical domain.

\section{Conclusion and Outlook}

In this work, it has been shown that interlocking time compensation in combination with a data\mbox{-}driven RLS model identification improves the prediction accuracy and, therefore, the performance of the FCS-MPC significantly compared to a baseline WB model which addresses varying parameter by offline identified LUTs.
Different data\mbox{-}driven model topologies have been compared against each other in terms of computational complexity and accuracy. Comprehensive experimental investigations proof the performance of the RLS\mbox{-}based FCS\mbox{-}MPC in steady\mbox{-}state and transients in the entire electrical and speed operation range. In particular, the presented approach is highly suitable for self\mbox{-}commissioning drive applications since accurate WB models are not available in this scenario.
In the future, it could be investigated how physical parameters can be extracted from the identified models in an appropriate way. These physical parameters could be used to calculate the maximum torque per current (MTPC) and maximum torque per voltage (MTPV) characteristics as well as the torque for a higher\mbox{-}level operating point control.

\bibliography{literatur_manually_reduced}

\end{document}